\documentclass{aa}  

\usepackage{graphicx}
\usepackage{url}
\usepackage{txfonts}
\usepackage{natbib}
\bibpunct{(}{)}{;}{a}{}{,} 
\usepackage{lscape}
\authorrunning{S. Kankkunen et al.}

\begin{document}

   \title{Active galactic nucleus time-variability analysis and its caveats}
   \author{S. Kankkunen \inst{1,2}, M. Tornikoski \inst{1}, T. Hovatta \inst{3,1}
          }

   \institute{Aalto University Metsähovi Radio Observatory, Metsähovintie 114, FI-02540 Kylmälä, Finland
              \\
              \email{sofia.kankkunen@aalto.fi}
                \and
            Aalto University Department of Electronics and Nanoengineering, PO Box 15500, 00076 Aalto, Finland
                \and
            Finnish Centre for Astronomy with ESO, FINCA, University of Turku, Turku, FI-20014 Finland
            }

   \date{Received April 30, 2024; accepted October 24, 2024}

 
  \abstract
   {}
   {In this study, we demonstrate some of the caveats in common statistical methods used for analysing astronomical variability timescales. We consider these issues specifically in the context of active galactic nuclei (AGNs) and use a more practical approach compared to mathematics literature, where the number of formulae may sometimes be overwhelming.}
   {We conducted a thorough literature review both on the statistical properties of light-curve data, specifically in the context of sampling effects, as well as on the methods used to analyse them. We simulated a wide range of data to test some of the known issues in AGN variability analysis as well as to investigate previously unknown or undocumented caveats.}
   {
   We discovered problems with some commonly used methods and confirmed how challenging it is to identify timescales from observed data. We find that interpolation of a light curve with biased sampling, specifically with bias towards flaring events, affects its measured power spectral density in a different manner than those of simulated light curves. We also find that an algorithm aiming to match the probability density function of a light curve has often been used incorrectly. These new issues appear to have been mostly overlooked and not necessarily addressed before, especially in astronomy literature. 
   }
   {}
   {}


   \maketitle
%

\section{Introduction}

Active galactic nuclei (AGNs) exhibit strong time variability in their flux densities. The extremity of the variability depends on the source, frequency domain, and various observational effects, such as Doppler boosting. The variability can be seen in a light curve as changes in its flux density, and generally interest is on the amplitudes and intervals of these changes. That is, one is typically interested in how often such changes occur, how dramatic they are, and whether they differ between observing periods.

The term `timescale' is usually associated with analysing time variability, although the meaning of timescale is ambiguous and varies from author to author. A timescale is often derived from the power spectral density (PSD) of a source. We discuss the PSD in Sect. \ref{PSDan} and how to obtain a descriptor called the periodogram for it in Sect. \ref{periodogram}. Analysing the PSD of a time series is very common in all signal-processing fields, which means that literature on the subject is vast and the methods are thoroughly documented. However, the information is spread across disciplines, which may complicate finding the relevant literature and then understanding the implications on AGN data analysis. 

Multiple authors have made great efforts in collecting, analysing, and improving time-series methods in the context of astronomical time-series analysis (e.g. \citealt{press1978flicker}; \citealt{scargle1979studies}; \citealt{uttley2002measuring}; \citealt{vaughan2003characterizing}; \citealt{vaughan2016false}). We refer to their work several times throughout this paper but also extend our literature review to other fields, specifically statistics literature. Some aspects of time-series analysis are often taken at face value, and we provide more information on them, hoping it will benefit future studies. 

We have written two articles jointly. In the first article, hereafter Paper I, we analysed the long-term radio variability of AGNs in the radio domain. We used up to 42 years of data observed in 37 GHz in the Aalto University Metsähovi Radio Observatory (MRO), and we attempted to find characteristic timescales for each source using the periodogram. The characteristic timescale we were interested in is one that can be seen in the periodogram as a bend frequency. We were only able to constrain this timescale for 11 out of 123 sources, and we believe a major reason to be the long timescales expected in the radio domain as well as sampling-induced effects. The timescales are not likely to be on the order of tens of years, but we believe their identification requires monitoring periods to be equivalent to several multiples of the timescale, depending on the sampling cadence. 

During the analysis process, we encountered some difficulties and unexpected results in our test simulations that we were initially unable to explain. We conducted a thorough literature review and performed more simulations in order to understand the root cause of the inconsistencies. The information we present here is a collection of both old information not necessarily widely known in the astronomy community as well as information we identified during the analysis in Paper I. We wish to share our findings, as they should aid in understanding some of the common pitfalls and misunderstandings in AGN variability analysis. We refrain from including in-depth mathematical descriptions and instead focus on the issues at a more practical level. 

In Sect. \ref{Astronomicaldata}, we discuss some basic concepts of astronomical data. We then continue to stochastic processes and how to model them in Sect. \ref{stochprocessmodel}. In Sect. \ref{methods}, we discuss the most commonly used and basic methods of AGN variability analysis. After this, we move to describing light-curve simulations and how they are used in model fitting in Sect. \ref{simulations}. Before summarising our review and analysis, we discuss characteristic timescales in more detail in Sect. \ref{Timescales}.

\section{Astronomical data} \label{Astronomicaldata}

Astronomical data is usually unevenly sampled, it has errors caused by observing instruments and environmental effects, and it may contain seasonal gaps. Ground-based astronomy, specifically of faint sources, is challenging due to the weather conditions altering the instrument detection limits. In the context of AGN studies, for blazars this is usually not a factor as they are highly beamed and thus bright sources. In the radio domain, for faint sources such as narrow-line Seyfert 1 galaxies \citep{lahteenmaki201737}, this is an important consideration when analysing the data. Here we focus on the analysis of blazars and other bright AGNs, which in addition to their high flux densities, are usually densely sampled, allowing for the use of specific time-series analysis tools, especially the use of the periodogram (Sect. \ref{periodogram}). The discussion in this article applies to all frequency domains, but some of the specific considerations are most relevant in cases where AGN variability is expected to be slow. 

\subsection{Light curve properties} 

\subsubsection{The probability density function}

An important feature of a time series is the probability density function (PDF). The PDF characterises the flux density distribution of the data points, that is with what probability can the data have a value between certain points. This means that we can for example identify the probability that a flux-density point is between 3 and 3.1 Jansky (Jy). The most common PDF is the Gaussian distribution, and it is also an assumption in many statistical analysis methods.  

The PDFs of AGN light curves cannot be purely Gaussian because they do not have negative values. Their non-Gaussianity also appears clear from plotting their PDFs, as their shapes are usually closer to a log-normal PDF though the specific shape depends on the frequency domain the observations are made in (e.g. \citealt{max2014method}; \citealt{liodakis2017bimodal}). Naturally, uneven sampling and observational errors complicate the analysis of the distribution. 

The log-normal distribution is common in nature, and it arises from multiplicative processes \citep{limpert2001log}. However,
\citet{scargle2020studies} challenges the necessity of multiplicative underlying processes to explain AGN PDFs and argues that the initial assumption of an 'either or' decision between Gaussian and log-normal distributions may be a false choice. We do not go into the details of this argument, but refer to the original article for more discussion. We discuss the PDF in the context of simulating light curves later in Sect. \ref{emm}.

\subsubsection{The power spectral density} \label{PSDan}

The power spectral density (PSD) of a time series is the modulus squared of the discrete Fourier transform (DFT) defined as (\citealt{deeming1975fourier})
\begin{equation} \label{PSD}
|F_N(\nu)|^2 = \left(\sum_{i=1}^{N} f(t_i) cos(2\pi \nu t_i)\right)^2 + \left(\sum_{i=1}^{N} f(t_i) sin(2\pi \nu t_i)\right)^2,
\end{equation}
where \(f(t_i)\) is a time series, N is the number of data points, t is the time of each observation, and \(\nu\) is the sampled frequency.

The PSD together with the process mean contains all information of the statistical properties of a stochastic process only when the process is Gaussian-Markov \citep{thorne2021statistical}. A Markov process is one whose present depends only on its previous state. The Ornstein-Uhlenbeck process discussed later in Sect. \ref{Orn} is one example of a Gaussian-Markov process. 

For non-Gaussian Markov processes, \citet{press1978flicker} shows how very different-looking light curves, generated by varying the duration of pulses as well as their frequency and amplitude, have the same PSD. Thus, it may be challenging to determine from visual analysis alone, whether the PSDs of two light curves differ from each other. We return to discussing PSD analysis later at multiple points, starting in Sect. \ref{spl}.

\subsection{Stationarity}

\begin{figure*}
    \centering
    \includegraphics[width=1\linewidth]{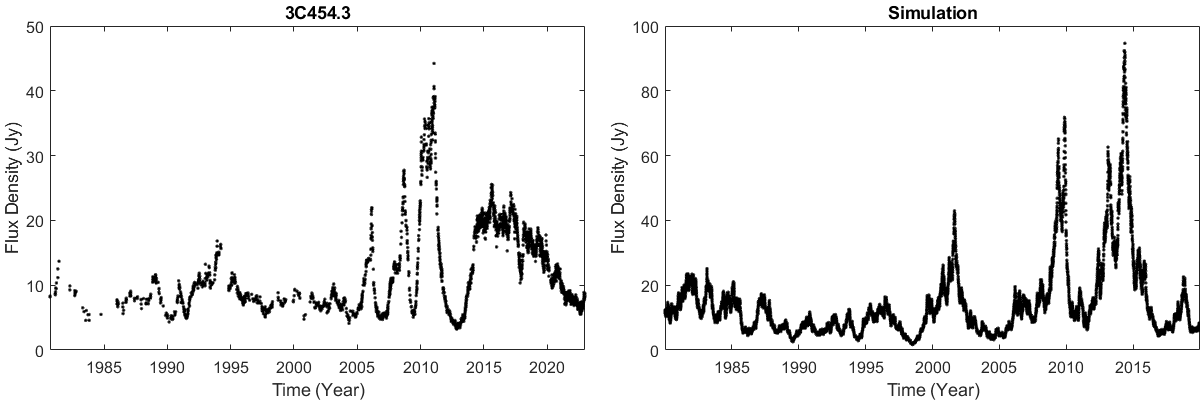}
    \caption{Comparison between observed data of 3C454.3 at 37 GHz (error bars not included) and simulated data on the right with similar modelling parameters obtained from the analysis of Paper I. The simulated data is evenly sampled, and the observed data has an uneven cadence.}
    \label{fig:3c4543}
\end{figure*}

For most statistical methods, the assumption is that the analysed data is stationary. This is self-explanatory, as defining a whole process from a discrete data segment requires the process to not change over time, especially not on observable timescales. A time series is stationary if its statistical properties, such as the mean, variance, and autocorrelation function, do not change over time. Unfortunately, this is not trivial to estimate from an AGN light curve due to statistical fluctuations. 

Source 3C454.3 is an example of a source that appears to visually change its behaviour over time as shown in the left-side plot of Fig. \ref{fig:3c4543} using 37 GHz data from MRO. On the right-side plot is a simulated light curve using similar PSD parameters (obtained in Paper I) and a log-normal PDF. We simulated 400 years of data and cut it to a segment of 40 years. The bend timescale is at 8.2 years, well within the observing period. The behaviour between the observed and simulated data is similar with extreme flare amplitudes towards the latter part of the observing period. The simulation is the product of a stationary model; The difficulties of determining stationarity in the context of AGN light curves are further discussed in \citet{vaughan2003characterizing}. 

A time series can be weakly or strongly stationary, and in research it is usually up to the author of the analysis to decide what they consider sufficient for their purposes. Weak or wide-sense stationarity indicates that some requirement of strong or strict stationarity is not fulfilled, while some other is. We return to the implications of stationarity later, as it is an important consideration in hypothesis testing.

\section{Stochastic processes and modelling them} \label{stochprocessmodel}

The astronomical data we observe are generated by stochastic processes. A stochastic process is a random process, where we are only able to derive certain statistical parameters of the process but not predict future values. Below we describe a few of the commonly used stochastic processes for modelling AGN variability behaviour and briefly discuss some of their caveats in the context of time-variability analysis. After that we move to describing what the colours of noise are, and then what the implications are for the shape of the PSD. For thorough reviews of stochastic processes and modelling them in the context of astronomy, we refer the reader to \citet{scargle1979studies} and \citet{scargle2020studies}.

While here we refer to the bending power law and simple power law as models, they are technically not models but descriptors for the distribution of power in the frequency domain. That is, they contain no information or assumptions of the underlying process generating similar PSD shapes.

The naming conventions of stochastic processes vary between disciplines and also within astronomical literature as noted by \citet{scargle2020studies}. Here we discuss models with names that are commonly seen in astronomy literature. 

\subsection{Gaussian processes}

Some stochastic processes are called Gaussian processes. The requirement is that a finite collection of data points generated by the stochastic process be jointly Gaussian. The Gaussian process is a generalisation of the multivariate normal distribution for infinite dimensions. Gaussian processes have been commonly used in astronomical studies to model light curves in the time domain \citep{kelly2014flexible}.

\subsubsection{Random walk}

Random walk is a stochastic process originally used to describe Brownian motion. Brownian motion is a Gaussian process that describes the path molecules take in a medium of liquid or gas, named after botanist Robert Brown (\citeyear{brown1828xxvii}). Today, random walk is also used to describe a variety of different processes, such as stock market changes and neuronal dynamics, which may not be Gaussian.

For a random walk, each position is the sum of the previous positions. In the most simple case this can be described as
\begin{equation}
y_t = y_{t - 1} + \epsilon_t,
\end{equation}
where t is time, y is the value of the random variable, and \(\epsilon\) is its error.

The equation tells us that each position of the random variable y at time t depends on its previous position only. 
When t \(\rightarrow\) \(\infty\), the random walk will take each position an infinite number of times, and thus the mean of the process is constant. However, its variance increases infinitely causing the process to be non-stationary. While the process is not stationary, the increments are both independent and stationary; that is, we should observe the same behaviour regardless of the initial value of observations.

\subsubsection{The Ornstein-Uhlenbeck process} \label{Orn}

The Ornstein-Uhlenbeck process \citep{uhlenbeck1930theory}, also called the damped random walk, is one of the more commonly used stochastic processes for modelling AGN behaviour in recent years (e.g. \citealt{kelly2009variations}). In contrast to the random walk, the Ornstein-Uhlenbeck process is a continuous-time process and a specific case of the Continuous Autoregressive process (CAR(1)) (\citealt{kelly2009variations}; for a review see e.g. \citealt{scargle2020studies}). The Ornstein-Uhlenbeck process is a modification of describing pure Brownian motion. The main difference is that the process variance is finite in the Ornstein-Uhlenbeck formulation whereas for pure Brownian motion it is infinite as discussed above. The Ornstein-Uhlenbeck process is mean-reverting -- that is, its value tends to its mean over time. It is also a stationary process as it has a well defined mean and variance.

The Ornstein-Uhlenbeck process has some caveats when used for modelling AGN noise processes. One of the issues is that it only allows for two free parameters, a bend frequency (described below in Sect. \ref{bpl}) and a normalisation \citep{kelly2014flexible}. This may for example cause some biases in the estimated characteristic timescales \citep{kozlowski2016degeneracy}. Another potential caveat with using the Ornstein-Uhlenbeck process is that it is a Gaussian-Markov process, that is its PDF is Gaussian. As we discussed previously, AGN light curves cannot be purely Gaussian due to their non-negative flux densities. 

\subsubsection{CARMA}

The Ornstein-Uhlenbeck process can be written as a first-order stochastic differential equation. \citet{kelly2014flexible} suggest adding higher order derivatives to the Ornstein-Uhlenbeck process to overcome the limitation of having only two free parameters. This model is called the continuous autoregressive moving average (CARMA) process, and it is a generalised version of the Ornstein-Uhlenbeck process but provides a more flexible PSD. Both the Ornstein-Uhlenbeck process and CARMA have an exponential autocorrelation function, which is desirable, as their likelihood functions can be calculated efficiently \citep{kelly2014flexible}. \cite{kelly2014flexible} show that CARMA can be used to model irregularly sampled stochastic and quasi-periodic light curves. Of course, with a higher number of parameters, one must then be careful when evaluating the significance of their results, as adding parameters will naturally risk overfitting the data.

\subsection{The simple power law and colours of noise} \label{spl}

The simple power law is defined as
\begin{equation} \label{spleq}
P(\nu) = \frac{1}{\nu^\beta},
\end{equation}
where the \(\nu\)'s are the probed frequencies and \(\beta\) is the PSD slope.

The simple power law is used for describing the different colours of noise. A process with a simple power-law PSD is not strictly stationary as its variance increases infinitely; that is it does not converge (e.g. \citealt{press1978flicker}). This is not feasible in a real process: In the context of AGNs, we should then see flare amplitudes growing infinitely over longer and longer temporal distances. 

While a simple power-law PSD cannot be produced by a real-world process over non-discrete observing windows, different slopes of the simple power law are used to define the so-called colours of noise.  Colours of noise are in practice simply definitions of the shapes of the PSD and do not hold physical meaning themselves. In real-world applications the processes are always band-limited, which relaxes their requirements for stationarity.

We can consider (the weak) stationarity of coloured noise in this way: If we have a data sample with a PSD that remains the same within that discrete observing window, each segment of the signal we take should have that same shape. In addition, any new observation of the same length of observing window should have the same PSD. 

\subsubsection{White noise} \label{whitenoise}

White noise is noise where all observed data points are uncorrelated in time. That is, the current position has no effect on the following position, and thus it has no memory of its past. White noise has finite mean and variance, and its PSD slope \(\beta\) is 0. Whether white noise is stationary or not depends on how it is defined, but by the usual definition it is. For our purposes it is not relevant to go into the mathematical details, but the reader should consider this possibility if reading further on the subject and seeing some apparent conflict in terminology. White noise can have a Gaussian distribution, but it is not a requirement.

\subsubsection{Red noise}

Another well-understood type of noise is red noise with a PSD slope \(\beta\) = 2. Red noise is obtained from integrating white noise and it has no correlation between increments, which means that in a time series the position of the next data point only depends on its current position. This can be thought of as its variance accumulating over every step, and hence the variance increases into infinity. The definition of red noise varies in astronomy: For example in the X-ray domain, all slopes \(\beta\) > 1 are often called red noise (e.g. \citealt{vaughan2003characterizing}). 

Pure red noise (\(\beta\) = 2) is produced by for example Brownian motion, hence it is also called Brown or Brownian noise (after Robert Brown, not to be confused with the colour). Noise produced by Brownian motion is by definition Gaussian and it has stationary increments (e.g. \citealt{hunt1956some}), but it is not a strictly stationary process as described earlier. For example, the PSD of random walk is red noise with \(\beta\) = 2 similarly to the damped random walk (CAR(1)), whose PSD is a Lorentzian.

\subsubsection{Flicker noise and other long-memory noise}

Flicker noise, also called \(1/f\) noise, as well as other coloured noise of slopes different from 0 or 2 continue to be poorly understood, as they cannot be obtained from integrating white noise. These processes require memory over long timescales, which is difficult to model mathematically \citep{keshner19821}.

Flicker noise is seen everywhere in nature, but it has proven to be very challenging to universally model these processes generating such noise (e.g. \citealt{press1978flicker}; \citealt{keshner19821}; \citealt{milotti20021} and references therein). On the other hand, it is debated whether there even is a universal explanation for flicker noise and that it would rather be a manifestation of different dynamics in different systems.

The definition of flicker noise is usually extended to include slopes between \(\beta\) = 0.5 to 1.5, as these slopes have been observed in electronics components (\citealt{milotti20021} and references therein). Flicker noise can be generated for instance by additive pulses of varying duration that have a power-law distribution \citep{halford1968general} or as a superposition of relaxation processes \citep{press1978flicker}. We discuss the generation mechanisms further in the context of characteristic timescales in Sect. \ref{Timescales}.

In AGN light curves the power-law part of the PSD is considered to be approximately red noise, but it typically varies between slopes \(\beta\) = 1 and \(\beta\) = 3 (e.g. \citealt{uttley2002measuring}; \citealt{max2014time}) and is dependent on the observing domain. In Paper I, the average slope of the power-law portion of the PSD for the 11 constrained sources was \(\beta\) = 2.3.

\subsection{The bending power law} \label{bpl}

Red noise and flicker noise are not strictly stationary due to their lack of convergence when f \(\rightarrow\) 0. This does not suggest that sources cannot have an apparent simple power-law slope over finite (band-limited) observing windows: The flattening may occur over frequencies whose inverse is longer than the monitoring period, and thus only the steep simple power-law portion of the PSD may be visible until more data is collected.

The bending power law has a power-law portion followed by a smooth bend to white noise in low frequencies. It can be defined as 
\begin{equation} \label{bpleq}
P(\nu) = \scalebox{1.2}{$\displaystyle \frac{1}{(1+\left(\frac{\nu}{x_b^{-1}}\right)^2)^\frac{\beta}{2}}$},
\end{equation}
where the \(\nu\)'s are the sampled frequencies, \(x_b\) is the inverse of the bend frequency, and \(\beta\) is the PSD slope \citep{uttley2002measuring}.

\begin{figure}
    \centering
    \includegraphics[width=1\linewidth]{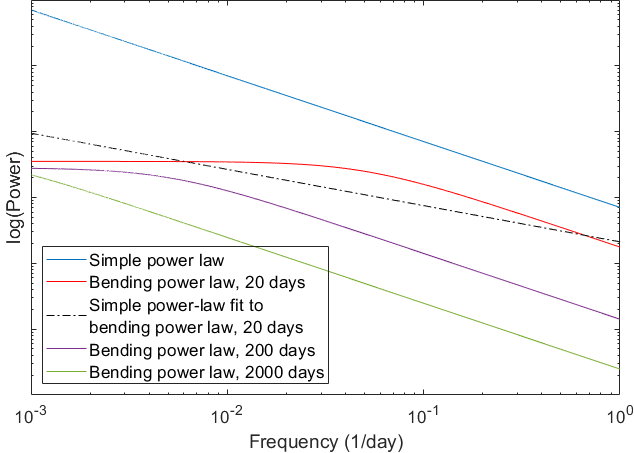}
    \caption{Plot showing four PSDs of varying parameters and with a slope of \(\beta\) = 2. In the blue is the simple power law, in red is the bending power law with a characteristic timescale of 20 days, in purple is the bending power law with a characteristic timescale of 200 days, and last in green is the bending power law with a characteristic timescale of 2000 days. The dashed grey line shows the principle of how a simple power law would fit to a bending power law. The powers (y-axis) are arbitrarily scaled.}
    \label{fig:bplspl}
\end{figure}

Figure \ref{fig:bplspl} shows the differences between the simple power law and the bending power law as well as how changing the timescale alters the PSD shape. As mentioned above, an important point to note is that the bending power law will equal the simple power law when the timescale is at a longer timescale than the observing window length (blue and green curves in Fig. \ref{fig:bplspl}). However, it is not quite this simple when analysing real-world data, but rather the timescale will need to be considerably longer than the observing window length for it to be guaranteed to not affect the fitting procedure. The considerations include uneven sampling, spectral leak as well as observational errors, which may all affect the PSDs in different ways depending on the PSD parameters. 

The plot also highlights the possible dangers in using a simple power law: If the bend frequency is already within the monitoring period length, using a simple power law to find the best fit for the PSD will result in a flatter estimate as shown in Fig. \ref{fig:bplspl} with the dashed line approximately fitted to the bending power law with parameters \(\beta\) = 2 and \(x_b\) = 20 days. Clearly, the dashed line would not be a good fit in this situation, but the principle is the same: Sparse sampling and observational errors distort the periodogram in a way that may lead to the bending power law resembling the simple power law more, especially if the characteristic timescale is closer to the length of the observing window.

The simple power law is still relatively commonly used in PSD analysis and it is not necessarily an issue per se. However, in addition to the potentially flat slope estimates, if the PSD of the source is not a simple power law, as is expected with long-enough monitoring periods, then using it as a reference model to identify deviations from the underlying noise process, such as quasiperiodicities, will result in false positives. \citet{vaughan2005simple} suggests that in such a case, using short enough segments of the light curve should allow this issue to be circumvented. Of course, if the quasiperiodicity occurs at such periods that it is within the flat portion of the PSD, other methods need to be considered. We discuss some other issues regarding the identification of quasiperiodicities from AGN light curves again in Sect. \ref{QPO}.

It is also possible that a bending power law is not enough to describe the PSD of a source. For example \citet{belloni1990variability} showed that the black-hole X-ray binary system Cyg X-1 includes two breaks in its PSD. Naturally, increasing the number of parameters poses its own constraints in analysis, and the risk of overfitting the data needs to be carefully assessed. For the bending and simple power laws, the fact that the bending power law becomes a simple power law in short-enough monitoring periods is convenient as it removes the necessity of comparing the model fits in relation to their different number of parameters. 

\section{Methods} \label{methods}

AGN variability analysis shares in principle all methods with other signal processing fields and even with stock market analysis as mentioned previously. Here we first describe the periodogram with some of its caveats we discovered during the analysis of Paper I, and then briefly a few of the other most basic methods that are used in AGN light-curve analysis.

\subsection{The periodogram} \label{periodogram}

The periodogram is an estimator of the underlying true power spectrum of a light curve and its use is widely covered also in astronomy literature (e.g. \citealt{uttley2002measuring}). The periodogram can be obtained by applying a normalisation to the modulus squared DFT (\citealt{deeming1975fourier}):

\begin{equation} \label{periodogrameq}
P(\nu) = \frac{2T}{\mu^2N^2}|F_N(\nu)|^2,
\end{equation}
where \(T\) = length of the time series, \(\mu\) = mean of the light curve, \(N\) = number of data points, and \(F_N\) = Discrete Fourier Transform. This normalisation allows the fractional root mean squared (RMS) variability to be determined \citep{uttley2002measuring}.

The mean of the periodogram approaches the true PSD as the length of the time series increases. However, the variance of the true power spectrum is exaggerated by the periodogram and the variance does not decrease with the number of data points increasing (e.g. \citealt{whittle1957curve}). This is especially important when analysing notable deviations from a null hypothesis of a featureless power spectrum. 

The periodogram probes only certain frequencies, which is important to remember when analysing light curves for any potential timescales. That is, for a light curve of length N, the periodogram (or more specifically the DFT) will search for sums of sine and cosine functions of periods T = N, N/2, N/3,.. exact periods depending on the chosen frequency grid and if oversampling is employed. Therefore, one must be careful when analysing for any specific timescales from the periodogram: If the timescale is not an exact probed Fourier frequency, the timescale will be shifted to adjacent bins. We discuss Fourier frequencies further in Sect. \ref{rednoise} and timescales in Sect. \ref{Timescales}. 

The periodogram is often binned logarithmically to reduce its variance \citep{papadakis1993improved}. The issue with logarithmic binning is that the low-frequency bins have a very small number of samples, thus making them much more erratic and difficult to fit compared to high frequencies \citep{uttley2002measuring}. However, if no binning is used or it is not done logarithmically, high frequencies will dominate as the contribution of low frequencies to any fitting procedure will be minimal. 

\subsubsection{Interpolation} \label{interp}

\begin{figure*}
    \centering
    \includegraphics[width=1\linewidth]{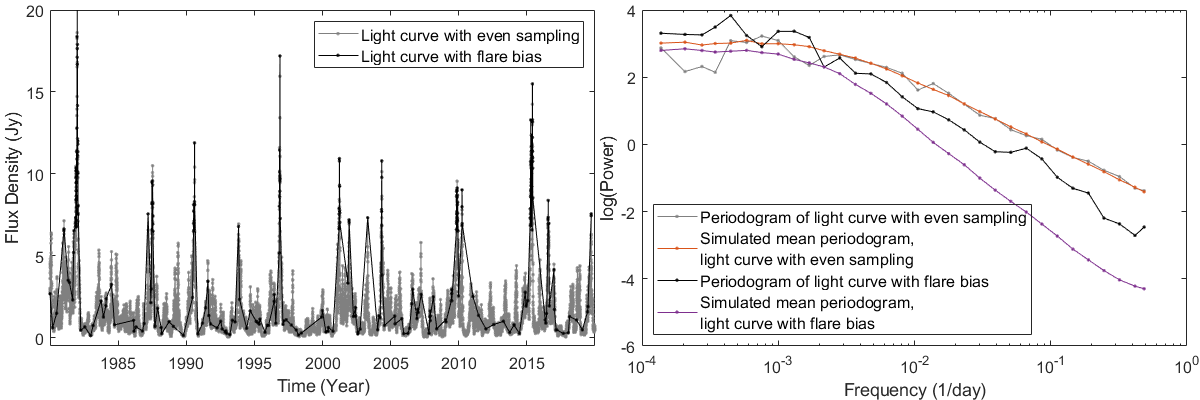}
    \caption{Comparison between the PSDs of an evenly sampled light curve and flare-biased light curve. Left-side plot: Light curve we generated using the bending power-law PSD with parameters \(\beta\) = 2 and \(x_b\) = 500. The grey light curve has been evenly sampled, and the black light curve has flare bias. Right-side plot: Obtained periodograms. The periodogram in grey shows the PSD of the light curve with even sampling. The periodogram in orange is the mean periodogram of 1000 realisations using the same parameters and even sampling. The black periodogram is the periodogram of the flare-biased light curve. The purple periodogram is the mean periodogram of 1000 realisations using the same parameters and replicated sampling. Because the replicated periods of denser sampling only randomly occur during flares in simulated light curves, the effect of interpolation is stronger than on the black periodogram.}
    \label{fig:flarebias}
\end{figure*}

The periodogram requires evenly sampled data; thus AGN light curves need to be interpolated. Interpolation may cause some issues if the data is very strongly biased towards flaring events, and if the time series contains sections of very sparse sampling. With strong flare bias, interpolation will affect the periodogram less than that of an unbiased sparsely sampled light curve. This occurs because interpolation smooths out short-timescale variations (reduces the power in high frequencies in the periodogram) causing the periodogram to become steeper. If the observed data instead contains many data points in high-activity states, the periodogram will contain more power in the high frequencies, thus decreasing the steepening effect of interpolation. This is an issue for model fitting as simulations that replicate the original sampling cannot reproduce the bias due to the randomness of a stochastic process: In our simulations the flares will occur randomly, and not necessarily during the periods of denser sampling. 

Figure \ref{fig:flarebias} shows an example of a periodogram constructed from a light curve with simulated flare bias and sparse sampling, and then conversely a periodogram constructed from a light curve with daily even sampling. We generated the flare-biased sampling by choosing data-point time stamps with pseudo-random Gaussian (integer) sampling, heavily favouring higher flux densities by dividing the time stamps of flux densities into two arrays based on their values. The simulation process is discussed later in Sect. \ref{psresp} and \ref{TK95}.

The black periodogram of the light curve with flare bias is steeper than the periodogram with even sampling (grey) due to interpolation. The steepening of the periodogram is caused by the flare-biased light curve having multiple gaps, which interpolation smooths out. However, the simulated mean periodogram using the sampling of the flare-biased light curve is clearly steeper. With flare bias, some portions of the light curve with high flux densities are over-represented: Interpolating between the points will hardly smooth these densely sampled flares resulting in more power on high frequencies. Because we are applying the same sampling to the simulated light curves used to construct the mean periodogram, the biased sampling will randomly occur during flares and randomly during periods of low flux densities. Therefore, the steepening of the periodogram due to interpolation will be stronger for the simulations, as many of the light curves used to construct the mean periodogram will have at least some strong flares with poor sampling. If there is less flare bias, the source light curve will be more similarly steepened to the mean periodogram of the simulations. 

The flare bias of MRO light curves is much less prominent than in our simulations. While the MRO light curves are thus less susceptible to these effects, it is a consideration and highlights the need for estimating the potential effects in the results gained from variability analysis of AGNs. 

\begin{figure*}
    \centering
    \includegraphics[width=1\linewidth]{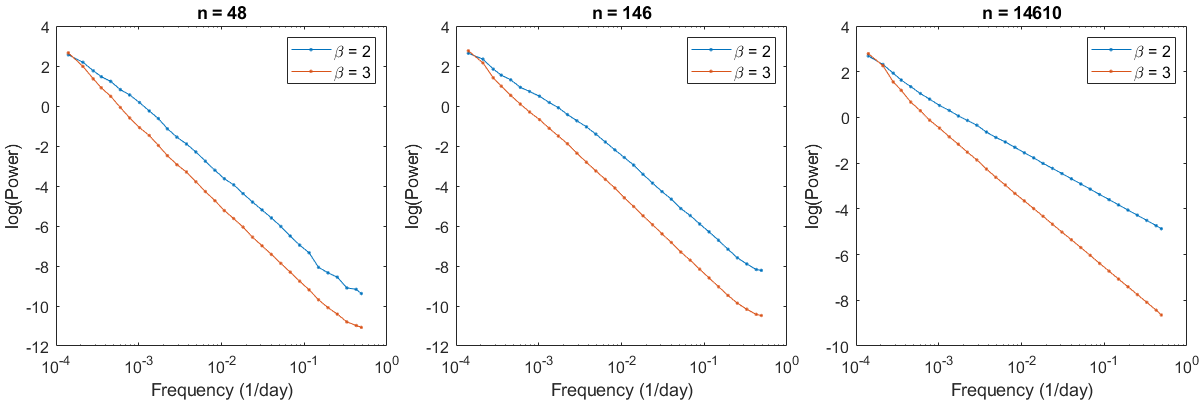}
    \caption{Mean periodograms plotted using simulated light curves constructed from two simple power laws with \(\beta\) = 2 and \(\beta\) = 3 with a varying number of data points. The first plot includes 48 data points, the second plot 146 data points, and the third plot is evenly sampled with the full 14610 data points.}
    \label{fig:sampling}
\end{figure*}

In Paper I, many sources also had a very large parameter space of good fits. We assumed this to be because interpolating sparsely sampled data results in similar periodogram shapes regardless of the model PSD used to create the light curves. Figure \ref{fig:sampling} shows how the difference between slopes \(\beta\) = 2 and \(\beta\) = 3 increases with denser sampling. The plots on the left and middle show how initially with sparse sampling the slopes are much closer to each other than with the densely sampled plot on the right. These effects are magnified with the bending power law, as it contains a second free parameter, as well as with the observational errors included. Here, the sampling is Gaussian and thus not completely representative of real sampling, which may include larger gaps and segments of denser sampling and further contribute to the number of good fits.

\subsubsection{Normalisation of the periodogram} \label{norm}

\begin{figure}
    \centering
    \includegraphics[width=1\linewidth]{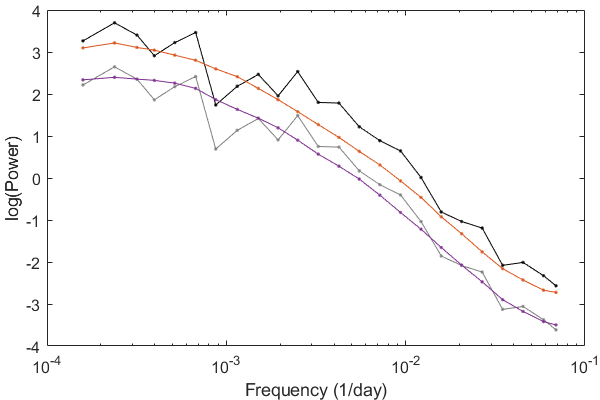}
    \caption{Examples of scaling the periodogram. The black and orange periodograms use the normalisation presented in Eq. \ref{periodogrameq}, whereas the grey and purple periodograms are shifted to have a mean of zero in logarithmic space.}
    \label{fig:zeromean}
\end{figure}

How the normalisation is done depends on the author, and especially on if the exact powers in the periodogram are of interest. When analysing timescales, the power axis can usually be chosen freely, and thus the choice of normalisation is less important. However, this is where it is relevant to consider the accuracy of the mean and variance of the observed light curve, if the light curves are shifted and scaled according to them. \citet{uttley2002measuring} consider this when averaging the light curve means in such a way, where an increased number of data points in one bin does not contribute to the mean more than a bin with fewer data points. However, as they also note, the mean of the light curve is vulnerable to variations inherent of a red noise process and adding uneven sampling may distort the statistical parameters further. 

We tested the normalisation of the periodogram with simply scaling the periodograms as described in Eq. \ref{periodogrameq}, and with an alternate strategy, where the periodograms were shifted to have a mean of zero in logarithmic space. This ensured that the position of a periodogram was independent of the statistical parameters obtained from the light curves. The motivation for testing this strategy came from initial simulations, where we noticed that using the light curve mean and variance for shifting and scaling the simulated light curves sometimes resulted in a slightly mis-matched power-axis position.

Figure \ref{fig:zeromean} shows periodograms with both the normalisation introduced in Eq. \ref{periodogrameq} as well as the shifted equivalents. Shifting the periodograms to have a mean of zero appears to bring the simulated mean periodogram closer to the periodogram of the mock light curve. The result is desirable as it removes some of the potential errors in the scaling of the simulated light curves. 

While shifting the periodograms to have a mean of zero was successful, it also had some issues, specifically due to the observational errors: Sources with flatter PSD slopes were not affected by this, but for sources with steep PSD slopes (\(\beta\) > 2) we were sometimes unable to set an upper limit for the slope. This is due to observational errors, which can limit the steepness of the slope; that is, adding Gaussian white noise to a smooth red-noise periodogram causes the high frequencies to flatten. This affects the non-shifted periodograms less because the power-axis position of the simulated mean periodogram will be changed even if the steepness of the slope is not dramatically altered. However, in the analysis of Paper I, we disregarded some fits that were clearly caused by observational errors as they were distinguishable as a sudden increase in fit significance in the steepest slopes (\(\beta\) > 3). 

\subsection{Briefly on other methods}

In the following sub-sections, we briefly describe the other methods that were tested for the analysis in Paper I. We conducted various simulations to confirm the caveats described in some of the original papers cited below.

\subsubsection{The Lomb-Scargle periodogram}

The Lomb-Scargle periodogram is a modification of the original formulation of the periodogram (\citealt{lomb1976least}; \citealt{scargle1982studies}). It is intended for unevenly sampled data and as such it has been widely used in AGN variability studies. The generalised Lomb-Scargle periodogram \citep{zechmeister2009generalised} is an improved modification of the original formulation, where observational errors are considered and the mean of the data need not be the mean of the fitted sine function. We tested the ability of the Lomb-Scargle periodogram to identify the correct PSDs, and confirmed that especially for longer periods, the generalised Lomb-Scargle periodogram was superior.

In the original formulation by \citet{scargle1982studies}, the false-alarm-probability (fap) is presented as a way to identify non-stationarities, such as quasiperiodicities. A sometimes overlooked aspect is that the fap is in fact only usable for stochastic processes generating white noise and for identifying a signal specifically from white noise (e.g. \citealt{vanderplas2018understanding}). Thus fap should not be used in the context of AGN variability analysis, where the data is red noise.

Most of the issues of the Lomb-Scargle periodogram are highlighted in \citet{vanderplas2018understanding}, where many common misconceptions over its superiority compared to using the ordinary periodogram are disproved. For example, it is often argued that the Lomb-Scargle periodogram does not suffer from aliasing (discussed in Sect. \ref{alias}), but \citeauthor{vanderplas2018understanding} shows this not to be true. 

\subsubsection{The structure function} \label{SF}

The structure function is in principle the time-domain version of the periodogram. Thus it should give similar information as the PSD. The structure function has been used for example to analyse phase and frequency instabilities in precision frequency sources (\citealt{rutman1978characterization}), and flicker noise remains an active field of study in electronics components (e.g. \citealt{schmid2018offset}). The first-order structure function is defined as (\citealt{simonetti1984small}; \citealt{hughes1992university})

\begin{equation} \label{sfeq}
    D^1(\tau) = \langle[S(t) - S(t + \tau)]^2\rangle,
\end{equation}
where S(t) is the signal flux density at time t and \(\tau\) is a time lag.

In the astronomy literature, the structure function is commonly described with having a flat short-timescale slope dominated by noise  \citep{hughes1992university}. This flat portion is then followed by a steeper slope, after which it turns over into an uncorrelated flat section. This is the same behaviour that we expect in a bending power law in the frequency domain. 

We simulated multiple sets of 100 light curves using the bending power law with \(\beta\) = 2 and varying the timescale location. We constructed both the periodogram and the structure function for them and visually estimated the bend-frequency location from the periodogram and the timescale from the structure function. The results from the structure function analysis show that the timescales are often close to a location \(x_b/2\), whereas in the periodogram the bends are close to \(x_b^{-1}\). This is a similar result that \citet{emmanoulopoulos2010use} obtained fitting a broken power law to the structure function. We did not consider all possible scenarios, specifically we did not use long time series with even sampling due to the computational cost. While we cannot draw any firm conclusions of a systematic offset until more simulations are done, the difference in the timescale obtained from the periodogram and structure function are known and reported in the astronomy literature (e.g. \citealt{tanihata2001variability}; \citealt{hovatta2007statistical}). The difference in the timescale location is often attributed to the structure function probing the rise and decay times of flares. From Eq. \ref{sfeq} we see that the structure function indeed is sensitive to differences in flux densities. The periodogram on the other hand deconstructs the light curve into sine (cosine) waves, where analysing a singular sine wave results in a peak in the periodogram, shown also later in the bottom left panel in Fig. \ref{fig:rednoiseleak}. If we analysed the sine wave with the structure function, the full period from start to finish would result in no difference in flux densities. Rather, the largest difference would come from the extremes of the sine wave, equating to half of the sine wave period. A similar explanation is provided in \citet{hovatta2007statistical}.

The slope \(\alpha\) of the structure function is also related to the PSD as \citep{emmanoulopoulos2010use}
\begin{equation}
    \alpha = \beta - 1.
\end{equation}

Similarly to the Lomb-Scargle periodogram, there are some misconceptions regarding the structure function. Its caveats are thoroughly discussed in \cite{emmanoulopoulos2010use}: For example, sampling affects the shape of the structure function significantly, and it cannot be used without simulations. The correlation between the PSD slope and the slope of the structure function depends on the light curve properties. That is, it needs to be stationary, have a mean of zero, have the frequency range \([0,\infty]\), and have a PSD slope of a power-law form with 1 < \(\beta\) < 3 \citep{emmanoulopoulos2010use}.

In addition to testing for the location of the characteristic timescale, we also generated structure functions for our simulated light curves and saw that densely sampled light-curve segments dominated them. This may warrant using some sort of a tapering method to ensure that the shape of the structure function is not biased towards areas of denser sampling. 

Because the structure function should give us the same information as the periodogram but the method contains more uncertainties, we opted to not use it in our final analysis in Paper I. 

\section{Simulations and model fitting} \label{simulations}

\subsection{The power spectral response method} \label{psresp}

We do not discuss the power spectral response (PSRESP) method in depth, but the reader is referred to the original article by \citet{uttley2002measuring}. The main principle is to use the original sampling of the source light curve and apply it to the simulated light curves to mimic sampling artefacts and spectral distortions. The goodness of fit is then estimated with a \(\chi^2\)-type distribution, \(\chi_{dist}^2\). 

In the PSRESP method, the \(\chi^2\)-value is obtained with the following equation:

\begin{equation}\label{chisq}
\chi_{dist}^2 = \sum_{\nu = \nu_{min}}^{\nu_{max}} \frac{(\overline{P_{sim}}(\nu)-P_{obs}(\nu))^2}{\Delta \overline{P_{sim}}(\nu)^2},
\end{equation}
where \(P_{obs}\) is the periodogram of the observed data, and \(\overline{P_{sim}}\) the mean of the simulated periodograms. The \( \chi_{dist}^2\) distribution is otherwise similar to the chi-square statistic, but it does not assume Gaussianity. \citet{uttley2002measuring} state that a major benefit of this approach is our ability to estimate the periodogram errors from the spread of the simulated periodograms.

The original formulation draws from an approximately Gaussian distribution to generate light curves (discussed later in Sect. \ref{TK95}). In our simulations, we did not observe issues with the ability of the PSRESP to identify the correct PSDs within error margins, even if the input light curve was highly non-Gaussian. The PSD and PDF of a light curve are connected through variance, where for a stationary zero-mean process, the integral of the PSD equals the process variance. 

\subsection{Spectral distortions}

\subsubsection{Red-noise leak} \label{rednoise}

\begin{figure*}
    \centering
    \includegraphics[width=1\linewidth]{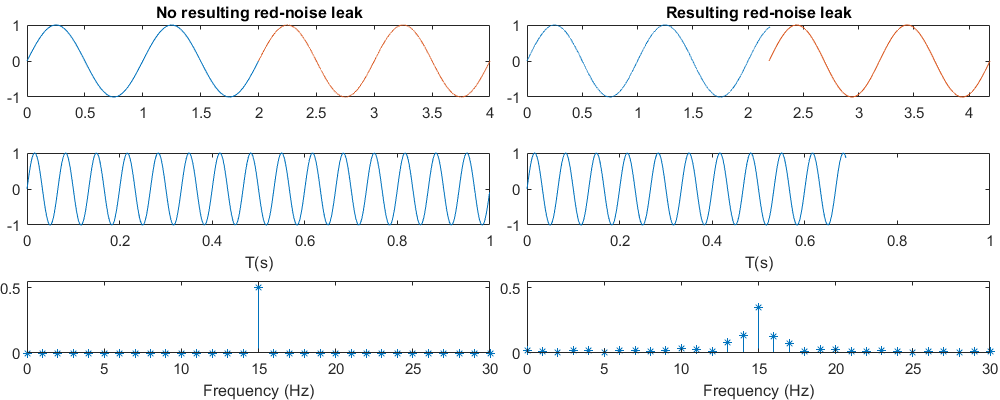}
    \caption{Plots illustrating the cause of red-noise leak in Fourier analysis. The top left-side plot shows a sine wave with two full periods resulting in no red-noise leak, depicted by the continuous sine wave transfer from blue to orange. The right-side plot shows a non-integer multiple of the period resulting in a discontinuity in the Fourier transform. The middle row shows sine waves with periods T = 1/(15 Hz), with the left-side plot having again full integer periods and the right-side plot conversely having a discontinuity. The bottom row shows the respective periodograms.}
    \label{fig:rednoiseleak}
\end{figure*}

\begin{figure}
    \centering
    \includegraphics[width=1\linewidth]{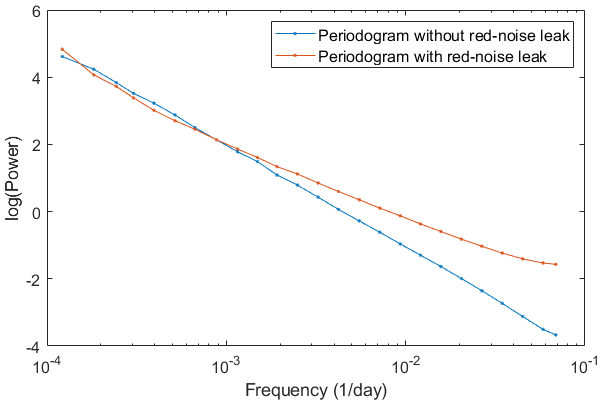}
    \caption{Comparison between periodograms with no red-noise leak and with red-noise leak present. The blue periodogram is of a light curve without red-noise leak, and the orange periodogram is of a light curve with red-noise leak. Both of the light curves were simulated using the same power-law slope.}
    \label{fig:rednoise}
\end{figure}

Red-noise leak is a well known issue considered in most astronomy literature where PSD analysis is conducted. It is caused by the finite observing window, and corrected for by initially simulating longer time series and then selecting a sub-section to represent the same finite observing window. We wish to expand on this explanation, as it is not usually discussed further in astronomy literature and hopefully the explanation provides some insight into the simulations. 

Red-noise leak occurs because the Fourier Transform of a rectangular window --- the window we are observing with --- is no longer rectangular, but a sinc function. This change in the window shape causes the frequencies to be spread to adjacent bins unless all deconstructed sine waves in the light curve have an integer multiplier of the number of periods in the observing window, resulting in exactly the observing window length. The Fourier transform essentially copies each individual sine wave, and if there is a discontinuity, spectral leak will occur. This can be visualised with a periodic signal in Fig. \ref{fig:rednoiseleak}. The top left-side plot shows a sine wave with two full periods. The orange part is essentially what the Fourier transform does; that is, it copies the input signal exactly as is. On the top right-side plot we see the discontinuity in this process, when the input sine wave does not have a full integer number of periods. The Fourier transform cannot generate a continuous pure sine wave from the original input signal. An impulse signal in the time domain is a constant in the frequency domain, which explains the consequent spread of the discontinuity. 

The middle plots of Fig. \ref{fig:rednoiseleak} correspondingly show two sine waves with period T = 1/(15 Hz) and below them are their respective periodograms. For the signal with an integer multiplier of periods (left-side plot), the peak of the power spectrum occurs exactly at 15 Hz as expected and there is no power in adjacent frequencies. Conversely, for the right-side plot with a discontinuity, the power of the signal is concentrated at 15 Hz, but some of the power is spread across neighbouring frequencies. 

Figure \ref{fig:rednoise} shows how red-noise leak is visible in the logarithmically plotted periodograms of noise containing multiple different frequencies. The blue graph without red-noise leak continues as a simple power law, but the orange plot with red-noise leak has a much flatter slope due to the spread of power into adjacent frequencies. Red-noise leak most prominently affects steeper power-law slopes \(\beta\) > 2. This is because light curves with steep PSD slopes have little power in high frequencies, and thus red-noise leak has a more significant impact on them.

In astronomy, we usually initially simulate a longer light curve and then cut it to the length of the observed time series to mimic the discontinuity. It is unclear how long the so-called red-noise length should be to sufficiently mimic this effect. Certainly, one can always choose 1000 times the length of the original time series, but this is usually unnecessary and computationally expensive. In Paper I, we simulated light curves with a length multiplier of 10. This appeared to be sufficient for our purposes and we did not notice a difference compared to the larger multipliers we tested (up to 10 000). 

\subsubsection{Aliasing} \label{alias}

Aliasing is caused by discrete data sampling, where it is most dominant with evenly sampled data \citep{deeming1975fourier}. Aliasing occurs when power on intermediate frequencies of continuous data is leaked to the measured frequencies if the sampling occurs at a lower rate than the Nyquist rate. Aliasing causes high frequencies of a PSD to flatten similarly to red-noise leak. As was discussed previously, the Lomb-Scargle periodogram is not immune to this spectral distortion despite using unevenly sampled data as input and requiring no interpolation. In fact, \citealt{vanderplas2018understanding} argues that using the Lomb-Scargle periodogram may be detrimental for the analysis, since the spectral leak cannot be properly characterised.

Aliasing mostly affects flatter PSD slopes \(\beta\) < 1.5 because they contain more power in short temporal intervals. Aliasing occurs when a continuous signal cannot be sufficiently sampled, and thus it is intuitive that noise with more power on those short intervals would then be more significantly aliased.

Aliasing cannot be eliminated from the data, but binning of the data helps reduce it. The aliasing effect can also be reproduced in simulations by simulating a finer grid of data, and then sampling from it according to the sampling pattern of the observed light curve. This ensures that there are higher frequencies present in the initial simulations. 

\subsection{Methods for removing red-noise leak}

\subsubsection{End-matching} \label{endm}

One way to reduce red-noise leak is to match the ends of the time series manually. We can force the start and end points of a light curve to be close to each others' values by choosing a sub-segment of the light curve where the start and end points match as closely as possible (e.g. \citealt{fougere1985accuracy}; \citealt{schreiber2000surrogate}). In general, the larger the difference between the values at the start and end of a time series, the larger the flattening caused by red noise leak in the high frequencies. In our simulations, it was sufficient to have the flux densities close to each other in value, but they did not need to be an exact match to reduce the effect of red-noise leak to be negligible. It is likely good practice to test the required matching in each individual case, depending on how accurate the PSD is desired to be. As \cite{schreiber2000surrogate} note, the smoothness of the time series relates to how well the ends need to match.

\subsubsection{Windowing}

\begin{figure*}
    \centering
    \includegraphics[width=1\linewidth]{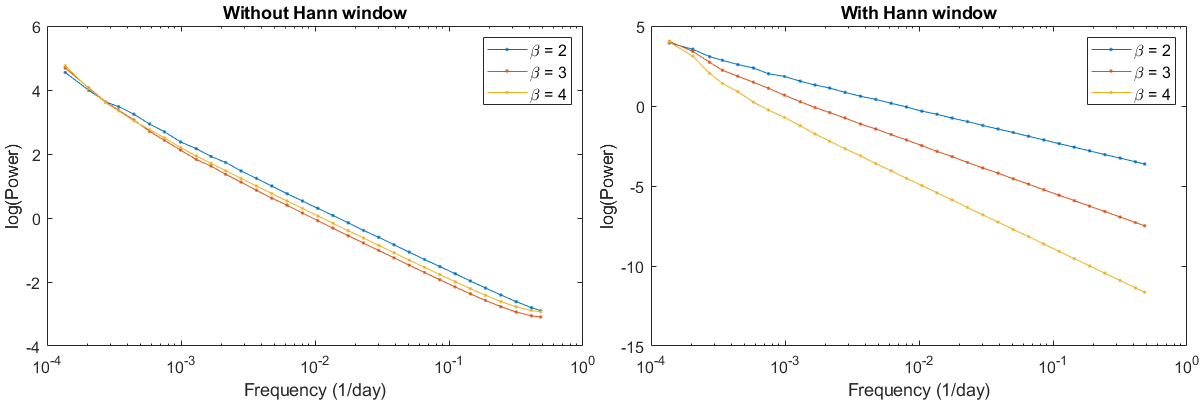}
    \caption{Effect of windowing on steep slopes in periodogram analysis. The left-side plot shows examples of periodograms with varying simple power-law slopes and red-noise leak present. On the right-side plot are the equivalent power-law slopes, but this time the same time series has been convolved with the Hann window.}
    \label{fig:Hann}
\end{figure*}

Red-noise leak can also be eliminated from the data fairly well by using a window function, but it is only applicable for evenly sampled data and is thus not possible with the Lomb-Scargle periodogram. \citet{max2014method} showed that a windowing function is necessary to avoid this flattening of the PSD in high frequencies unless some other method of reducing the leak is used. While technically the flattening will be imprinted on the simulated data when using the PSRESP method, finding the upper limits for the steepness of the PSD slope will be difficult. Figure \ref{fig:Hann} illustrates the issue. For the left-side plot, no windowing function is used and all periodograms with \(\beta\) = 2, 3, and 4 appear to have almost the same slope. In the right-side plot the slopes are clearly different from each other, and this is because the simulated light curves are convolved with a Hann window. What the windowing function in principle does is to weigh the discontinuous portion (both ends of the light curve) in such a manner that it becomes approximately continuous. This transform is naturally not perfect. 

The window function works similarly to end-matching in the sense that the start and end of the data is brought to zero. Windowing is not always unproblematic though. An issue rises, when there is relevant information at the beginning and at end of the observed time series because the windowing function smears this information out. When analysing stochastic noise processes and attempting to find a characteristic timescale, this is a possible issue because the low frequencies are smeared out. A priori it is difficult to know, whether the characteristic timescale is sufficiently within the observing period not to be affected by the windowing function. 

For a comprehensive review of windowing functions in the context of discrete Fourier transforms, we refer the reader to 
\citet{heinzel2002spectrum}.

\subsection{Simulated time series}

\subsubsection{The Timmer and Koenig method} \label{TK95}

The most commonly used method for time series simulations in the astronomy literature is the algorithm introduced by \citet{timmer1995generating}, hereafter TK95. The algorithm randomises both the phase and amplitude of the light curves. The idea is that the algorithm allows any realisation of the process to be generated.

\begin{figure*}
    \centering
    \includegraphics[width=1\linewidth]{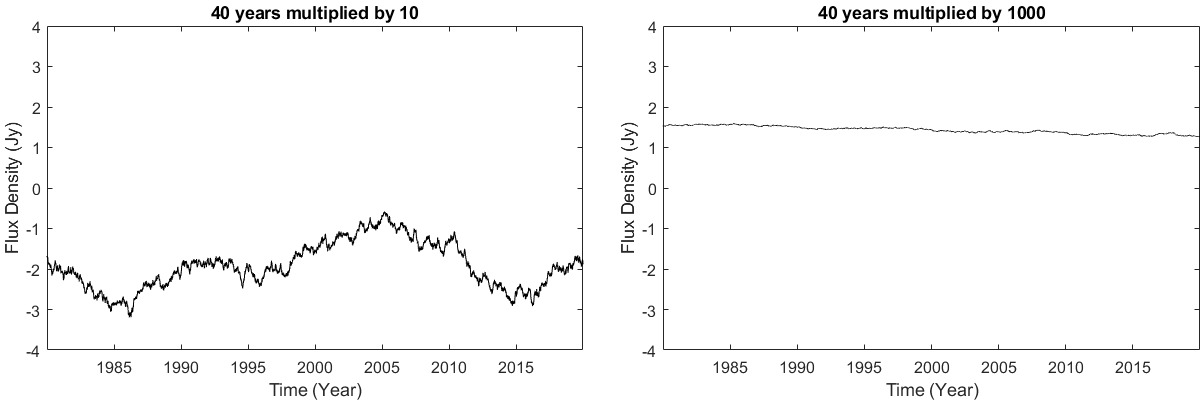}
    \caption{Comparison between simulated light curves when varying the simulation length in the TK95 algorithm. The left-side light curve is a 40-year segment from a 400 year-long simulated light curve using the TK95 algorithm and slope \(\beta = 2\). The right-side light curve has the same parameters but is cut from a 40 000-year long light curve.}
    \label{fig:TK}
\end{figure*}

Something that needs to be considered, when simulating light curves with the TK95 algorithm, is that the variance does not depend on the number of data points. This is especially important when simulating longer initial light curves in order to apply red-noise leak to the data. That is, if one simulates red-noise leak by increasing the length of the observing period by any factor, then the variance in the cut segment will be reduced according to that factor. Figure \ref{fig:TK} demonstrates this with using ten and 1000 times longer time series for the initial simulation, after which the data is cut to a 40-year segment. This is not an issue if the data is scaled but is nevertheless a consideration. 

\begin{figure*}
    \centering
    \includegraphics[width=1\linewidth]{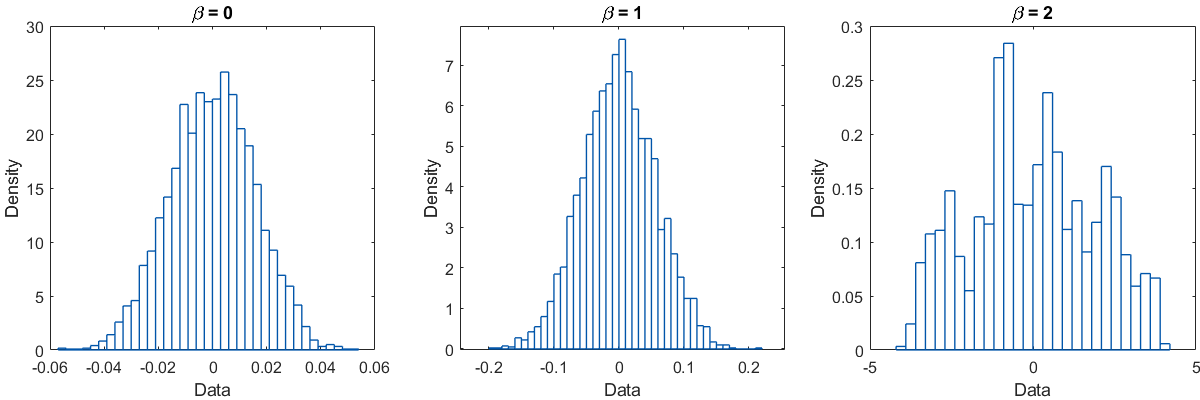}
    \caption{Distributions of flux densities from simulated data using the TK95 method with the simple power law and the slopes \(\beta\) = 0, \(\beta\) = 1, and \(\beta\) = 2, respectively.}
    \label{fig:TK_distr}
\end{figure*}

The Gaussianity of the light curve PDFs created by the TK95 method should also be considered carefully. \citet{morris2019deviations} showed that steeper PSD-slope values saw a decrease in the Gaussianity of the light curves. We saw this same effect in our simulations shown in Fig. \ref{fig:TK_distr}, where using a simple power law with either \(\beta = 0\) or \(\beta = 1\) gives clearly Gaussian-distributed data, but the Gaussianity is much weaker when \(\beta\) = 2. This is a serious consideration in analysis, where TK95 data is used as is with the assumption of its Gaussianity. 

\subsubsection{The Emmanoulopoulos algorithm} \label{emm}

\begin{figure*}
    \centering
    \includegraphics[width=1\linewidth]{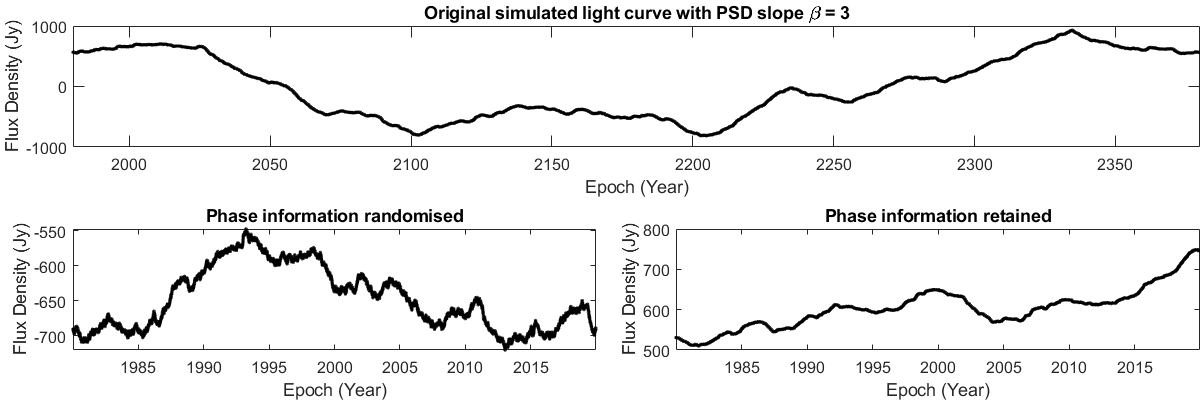}
    \caption{Plots illustrating how randomising the phase information affects the resulting light curve in the EMP13 algorithm. The top row is the entire simulated light curve with slope \(\beta = 3\) using the TK95 algorithm. The bottom left-side plot shows the light curve obtained from a 40-year segment of the entire light curve, which has been run through the iterative step of the EMP13 algorithm. The right-side plot shows the result of the same iterative step when the entire light curve has gone through it, and only after the iterations has it been cut to a 40-year segment. The PDF used for the EMP13 algorithm was Gaussian with a mean of zero and a standard deviation of one.}
    \label{fig:EMM13}
\end{figure*}

\begin{figure}
    \centering
    \includegraphics[width=1\linewidth]{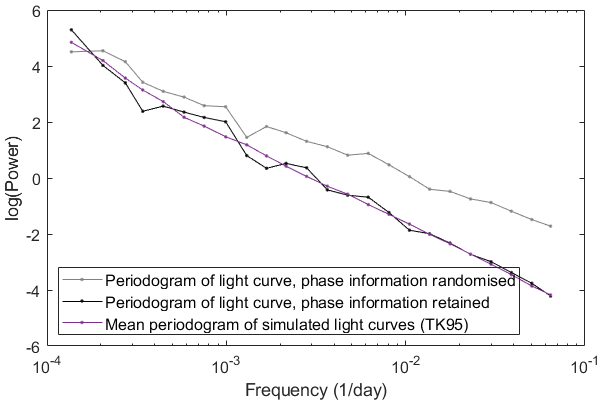}
    \caption{Resulting periodograms when using a window function on light curves with phase information randomised and with phase information retained. The light curves have been convolved with the Hann window to show the effect of phase randomisation in the iterative step of the EMP13 algorithm. The grey periodogram of the phase-randomised light curve is not affected by the Hann window, and thus it remains flat, similar to the left-side plot of Fig. \ref{fig:Hann}. The black periodogram, with phase information retained, is affected in the same way as the simulations made with the TK95 algorithm. With the TK95 algorithm, the phases are retained, as there is no iterative step.}
    \label{fig:Hannemm}
\end{figure}

An improved simulation algorithm by \citet{emmanoulopoulos2013generating}, hereafter EMP13, for generating simulated PSDs with a known PDF has been widely adopted in the astronomy community. The method uses the TK95 algorithm for its input data. 

The method is based on the Improved Amplitude Adjusted Fourier–Transform \citep{schreiber2000surrogate} (IAAFT). The method is a part of a group of methods intended for creating simulated data, also called surrogate data. There are differences between the EMP13 and IAAFT method, mainly that in the EMP13 formulation, the TK95 data is used as a basis for the surrogate time series and the original amplitude data is not simply shuffled, but drawn pseudo-randomly from a model distribution (usually created based on the observed data). 

The EMP13 method has a caveat, which has often been overlooked. In EMP13 chapter 2.3, it is suggested that if phase information is not required, the entire light curve can be simulated and then cut to the wanted length before the iterative (and thus the most time-consuming) step preserving the red-noise leak. However, this presents an issue: The iterative step in generating surrogate data attempts to end-match the light curve. More accurately, the jump between the first and last point of the shorter segment is spread through time in the iterative step because phase information is not preserved. With the generation of a Gaussian-distributed light curve in TK95, the phases are randomised but cutting the time series short for simulating red-noise leak preserves the same randomisation. If this time series is then used for the EMP13 iterative step, the phase information is again randomised. The red-noise leak will thus be imprinted on the PSD in such a manner that windowing methods used for reducing red-noise leak do not work. Figure \ref{fig:EMM13} demonstrates this with the top row showing the entire TK95-simulated light curve with simple power law and slope \(\beta\) = 3. The bottom left-side plot shows the resulting light curve when the phase information is randomised, that is, when the light curve is cut to a shorter segment before the iterative step (see the end-matched flux densities). The right-side plot demonstrates the result if the entire long light curve is used in the iterative step and cut to the desired length only after it. Visually, it is clear that the left-side plot contains more high frequencies than the original light curve or the right-side plot. 

This is the same issue that was discussed in Sect. \ref{rednoise} and it is demonstrated in Fig. \ref{fig:Hannemm}. When a light curve simulated with a steep slope (here \(\beta\) = 3) is convolved with the Hann window, the windowing function will correct for red-noise leak if the phase information is retained. If the phase information is randomised, as it is in the iterative step of the EMP13 algorithm, then the light curve convolved with the Hann window will result in a flatter periodogram. This occurs because the Hann window cannot recognise the presence of red-noise leak if the original phase information is not retained. Because red-noise leak starts to affect the periodogram when \(\beta\) > 2, this may not be a serious issue unless steep PSD slopes are analysed. The phase randomisation does not affect the results of model fitting as severely if no windowing functions are used, but it will reduce our ability of constraining the correct slope, and especially our ability to constrain an upper limit for it as explained before. 

Depending on application, one could potentially implement end matching (Sect. \ref{endm}) for the input light curve in order to avoid the computationally heavy simulation of a long light curve. That is, one could end match the original light curve and base the surrogate iterative step on this sub-segment and simply use a simulated light curve of the same length for the iterative step. The process is described in \cite{schreiber2000surrogate}. 

The original IAAFT method and thus the EMP13 method also suffer from difficulties with matching the PDF and PSD exactly to the wanted parameters \citep{maiwald2008surrogate}. This may be a similar issue noticed by \citet{morris2019deviations}, where it was shown that the TK95 method gives a poorly Gaussian PDF for steeper PSD slopes. It is likely inherent to the PSD that not all PDFs are possible for given models and model parameters. \citet{maiwald2008surrogate} show that following the final iterative step in the IAAFT method ends with a perfectly matching PSD but a less perfect PDF match. Stopping the algorithm before the final step in the iteration results in a perfect PDF match but a less perfect PSD match. They suggest that depending on the analysis, the researchers need to choose which information is more crucial. This information may also warrant specific care for shifting and scaling data as dramatic changes in the PDF may have unexpected effects on the PSD. 

\section{Characteristic timescales} \label{Timescales}

We discussed timescales briefly in the introduction but expand on them to complement the analysis in Paper I. 

Since the publication of the articles by \citet{uttley2002measuring} and \citet{vaughan2003characterizing}, it appears to have become more widely known in the astronomy community that a characteristic timescale seen in the PSD is a fundamental requirement of a real stochastic noise process and thus should also be expected in the PSDs of AGNs. The characteristic timescale in question will show as a bend from a steep power-law part to a flat white-noise portion in low frequencies, similarly to the PSD given by the bending power law. \citet{press1978flicker} discusses this bend frequency, but states that it may be at such a low frequency that it need not be considered. Indeed, there is discussion on when and if the timescale will be observable across disciplines (e.g. \citealt{milotti20021} and references therein), but defining a short and for that matter long monitoring period is challenging.  

The bend frequency has been recovered in the PSDs of some sources especially in frequency domains characterised by fast variability, specifically the X-ray domain. \citet{edelson1999cutoff} found a bend in the PSD of Seyfert I galaxy NGC 3516 at their estimate of approximately 1 month. \citet{uttley2002measuring} continued this work with three additional sources for which they found a flattening in the low frequencies for three of them, including for NGC 3516. In Paper I, using data streams that were up to 42 years long, we were able to identify that 11 out of 123 analysed sources in 37 GHz appeared to have a timescale visible in their PSDs. Similar results for a smaller sample of sources in 14.5 GHz were obtained by \citet{park2017long}.

\subsection{Characteristic timescale relationship with physics of AGNs} \label{flaredur}

In Paper I, we found a preliminary correlation between the characteristic timescale and the maximum time a knot is visible in 43 GHz very long baseline interferometry (VLBI) images. As knots are known to be correlated with the flares in the radio domain (e.g. \citealt{turler2000modelling}; \citealt{savolainen2002connections}; \citealt{lindfors2006synchrotron}), this naturally suggests that the flare duration is somehow connected to the characteristic timescale. We believe this to be logical, as power-law noise can be generated with pulsations of varying duration where the duration correlates with the bend frequency (e.g. \citealt{halford1968general}; \citealt{press1978flicker}). 

One way to think of the characteristic-timescale relation to the pulse duration is to consider a pulse followed by another pulse. As long as the first pulse remains non-zero, it will affect the second pulse by increasing the overall variance of the process. In the PSD, in principle, as long as the first pulse affects the following multiple pulses, the PSD slope will continue as a simple power law. When the initial pulse has fully decayed, it causes the simple power law to flatten, as a discontinuity in the power-law relationship would break down the increase of power and variance. We do not expect the pulses to have equal durations, and thus the bend frequency should refer to the longest pulse duration. \citet{press1978flicker} states that the continuing of the power-law slope would require a power-law relation between the pulse durations. 

Our description of the process may be too simplistic, but it is perhaps helpful in creating some sort of an intuition of how the PSD works. \citet{keshner19821} provides good discussion on the topic of coloured noise and its memory regarding different slope values.

\begin{figure}
    \centering
    \includegraphics[scale=0.6]{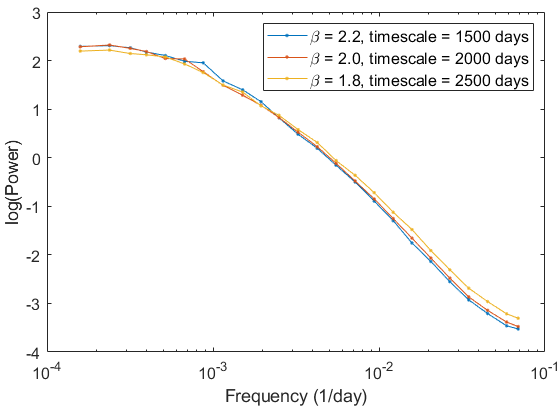}
    \caption{Mean periodograms of sparsely sampled light curves showing how small the differences are between the different parameter combinations.}
    \label{fig:betaxb}
\end{figure}

The identification of a characteristic timescale is not simple as was shown in Paper I: The outcome is always a large parameter space of good fits. In our analysis, we noticed that the characteristic timescale and the PSD slope had a clear correlation. We believe this to be related to the fitting of a model with two freely varying parameters, especially in sparsely sampled data. As was shown already in Sect. \ref{interp} with a simple power law, the cadence of the observations affect the periodogram significantly, and poor sampling hinders our ability to constrain a best fit. 

\citet{park2017long} discuss these results, that is, the correlation between the timescale and the slope, from a more physical perspective. Figure \ref{fig:betaxb} shows how the shapes of the periodograms evolve according to the parameter combinations when using heavily gapped data and a bending power law. We chose these combinations based on the results from Paper I, where we often saw these types of correlations between the slope and timescale. Already visually it is very difficult to differentiate these bending power laws from each other, and even with evenly sampled data, the differences are not very dramatic. Since it is clear that adding more model parameters generates better fits, and especially because of the uncertainties caused by sampling and observational errors, we understand the correlation between slope \(\beta\) and the bend frequency (timescale\(^{-1}\)) as their consequence.

\subsection{Identifying a quasiperiodicity} \label{QPO}

It is sometimes unclear what authors refer to as a quasiperiodicity. It is our understanding that quasiperiodicity suggests a deviation from the underlying noise process(es), and it should be visible as a deviation from the expected PSD. Some authors possibly consider this quasiperiodicity to simply be inherent of the noise process and use it to describe the rate of variability they derive from visual inspection of the light curve. However, as such it should then not be separable from the source PSD. \citet{press1978flicker} notes that to the eye, flicker noise appears to be periodic, which is an important consideration when attempting to perform visual analysis of a red noise process. That is, it is an inherent feature of coloured noise to contain such changes, which may appear to have a periodic nature.

Quasiperiodicity usually refers to a periodic component that does not repeat for the entire monitoring period. If there indeed exists a periodic 'signal' observable in AGN light curves, it should be additional to the inherent stochastic noise process, and thus we should see a separate component in the PSD. This is also how quasiperiodicities are usually probed from light curves, but determining the significance of such deviations is notoriously difficult: \citet{vaughan2005simple} and \citet{vaughan2016false} give prescriptions for how to analyse for quasiperiodicities in red noise. They assume a simple power law, which complicates things for long monitoring periods. 

There is also a concern in how to detect quasiperiodicities from a time series which includes red-noise leak, that is all AGN light curves. A quasiperiodicity, when at an intermediate frequency (that is, not at a multiple of the monitoring period length), will cause spectral leak. If the time series includes a quasiperiodicity, reconstructing a power spectrum becomes more difficult: As explained in Sect. \ref{rednoise}, the time series needs to be approximately continuous at the first and last data points. However, \citet{theiler1993detecting} show that if the time series includes a quasiperiodicity, also this periodicity needs to have an integer number of periods, which should also be an integer fraction of the time series length. Otherwise, its power will be spread to adjacent frequency bins and in this situation windowing the time series will not help.

\section{Summary}

Analysing AGN variability is convenient in the sense that the same signal-processing methods are used in most time-series analysis fields, and thus literature on the topic is vast. Nevertheless, or maybe for that reason, it is sometimes difficult to find exact information on why these methods work as they do and what their caveats are. 

A severe complication in the analysis of astronomical time series is of course that the test environment cannot be controlled. Therefore, it is important to carefully assess the benefits of using certain methods and whether they provide adequate results for each application. Unfortunately, it is not always trivial to find the right information, and even in statistics literature, there are differing opinions on some concepts. 

Our main conclusions are as follows:
\begin{enumerate}
    \item Analysis of AGN time variability requires careful consideration of how the methods work and which models are sensible for hypothesis testing. Even if a method has been widely used in a certain way, it is a good idea to test it with simulations in order to see if the results are as expected. 
    \item Surprising factors can affect the analysis, such as decisions on light-curve or periodogram bin size, whether to use windowing functions, which method to simulate the data with, how the data is normalised, how long the monitoring period is, and so on. Effects of these decisions on the analysis should be assessed, especially if very exact results are given.  
    \item Usually, the original formulation already gives restrictions on the use of the method or algorithm. For example, \citet{emmanoulopoulos2013generating} clearly specify that phase information is lost when cutting the light curve short before the iterative step, but what this actually means has not been well understood. Similarly, \citet{scargle1982studies} specifically discussed white noise and identifying signals from white noise. Regardless, fap has been used on red-noise PSDs.  
    \item While it is necessary to understand the used methods well, it is likely unnecessary to use overly complicated models for modelling AGN variability, especially without any physical a priori reasoning. 
    \item A characteristic timescale can refer to different things depending on the author, and its meaning is an important consideration when discussing AGN variability.
    \item Quasiperiodicities are too often proposed to explain AGN variability, even if the data  were consistent with noise. Care should be taken when considering what a quasiperiodicity is and how it may manifest in the data. 
\end{enumerate}

We have identified several caveats in the variability analysis of AGNs and described our findings in detail. The provided explanations and clarifications in methods should help other researchers to refine their approach in analysing astronomical time series.

\begin{acknowledgements}
    The authors would like to thank Dr. Kari Nilsson for his valuable comments on the draft. S.K. was supported by Jenny and Antti Wihuri Foundation, Väisälä Fund and Academy of Finland project 320085. T.H. was supported by Academy of Finland projects 317383, 320085, 322535, and 345899. This publication makes use of publicly available data from the Metsähovi Radio Observatory ({https://www.metsahovi.fi/opendata/}), operated by Aalto University in Finland. 
\end{acknowledgements}

%
%

\bibliographystyle{aa} 
\bibliography{sources} 

\end{document}